\documentclass[preprint,a4paper]{elsarticle}
\usepackage{amsmath,amsfonts,euscript,amssymb}
\usepackage{epsfig}
\usepackage{graphicx}

\newcommand{\be}{\begin{equation}}
\newcommand{\ee}{\end{equation}}
\newcommand{\bea}{\begin{eqnarray}}
\newcommand{\eea}{\end{eqnarray}}
\newcommand{\dd}{\mbox{d}} 

\begin{document}

\begin{frontmatter}

\title{Radiative decays of radially excited mesons 
${\pi^0}'$, ${\rho^0}'$, $\omega'$ in NJL model}

\author[bltp]{A.B.~Arbuzov\corref{cor1}}
\ead{arbuzov@theor.jinr.ru}
\author[bltp]{E.A.~Kuraev}
\author[bltp]{M.K.~Volkov}

\cortext[cor1]{Corresponding author}

\address[bltp]{Bogoliubov Laboratory of Theoretical Physics,
Joint Institute for Nuclear Research, 141980 Dubna, Russia}

\begin{abstract}
Radiative decays $\pi^0({\pi^0}')\to\gamma+\gamma$, 
${\pi^0}'\to\rho^0(\omega)+\gamma$, 
${\rho^0}'(\omega')\to\pi^0+\gamma$, and
${\rho^0}'(\omega')\to{\pi^0}'+\gamma$
are considered in the framework of the non-local SU(2)$\times$SU(2) NJL model.
Radially excited mesons are described with the help of polynomial form factor 
$f({k^\bot}^2)$, where $k^\bot$ is the quark momentum transverse
to the external meson momentum. In spite of mixing of the ground and excited
meson states in this model, the decay widths of $\pi^0\to\gamma+\gamma$
and $\rho^0(\omega)\to\pi^0+\gamma$ are found to be in good agreement
with experimental data as in the standard local NJL model.
Our predictions for decay widths of radially excited mesons can be
verified in future experiments.  
\end{abstract}


\end{frontmatter}

High luminosity modern electron-positron accelerators with CMS energy 
of about several GeV ({\it e.g.} BEPC-II (Beijing), VEPP-2000 (Novosibirsk),
DA$\Phi$NE (Frascati)) allow studying properties of different mesons
with masses up to 3~GeV, including radially excited meson states.    
Let us note that the standard local quark Nambu-Jona-Lasinio (NJL) model allows to 
describe the low energy meson 
physics in a satisfactory agreement with the 
experiment~\cite{Ebert:1982pk,Volkov:1984kq,Ebert:1985kz,Volkov:1986zb,
Vogl:1991qt,Klevansky:1992qe,
Volkov:1993jw,Ebert:1994mf,Volkov:2006vq}. 
In these papers, the properties of the ground meson states were considered.
A  non-local version of the  NJL  model with a polynomial form factor
was used in order to describe also the first radial excitations of 
scalar pseudoscalar and vector mesons~\cite{Volkov:2006vq,Volkov:1996br,Volkov:1996fk,Volkov:1997dd,Volkov:1999yi}.
The mass spectrum and strong interactions of mesons were studied there.
Here we In this paper we consider radiative decays of ${\pi^0}'(1300)$,
${\rho^0}'(1450)$ and $\omega'(1420)$ which can measured in the current
and future experiments~\footnote{
Recently the two-photon decay widths of $\pi^0$ and ${\pi^0}'$ and the 
cross sections of these meson production in $e^+e^-$ collisions
were calculated in Ref.~\cite{Kuraev:2009uh}. Unfortunately, 
some technical  mistakes were made there. They will be corrected in the present 
paper (see also v.2 of {\tt arXiv:0908.1628}).}.

The form factor of the most simple polynomial type is chosen:
\bea
f_{\pi,\rho}({k^\bot}^2) &=& c_{\pi,\rho}f({k^\bot}^2), 
\\ \nonumber 
f({k^\bot}^2) &=& (1-d |{k^\bot}^2|) \Theta(\Lambda^2-|{k^\bot}^2|),
\qquad
{k^\bot} = k - \frac{(kp)p}{p^2},
\eea
where $k$ and $p$ are the quark and meson four-momenta, respectively.
In the rest frame of the
external meson ${k^\bot}=\{0,\vec{k}\}$ and ${k^\bot}^2=-\vec{k}^2$.
The parameter $c_{\pi,\rho}$ defines only the meson masses and can be omitted in
the description of meson interactions. 
The cut-off parameter is taken to be $\Lambda=1.03$~GeV. 

In our model the slope parameter $d$ is chosen from the condition so that
the excited scalar meson states do not influence the value of the quark condensate,
{\it i.e.} do not change the constituent quark mass $m_u=m_d=280$~MeV 
(see Refs.~\cite{Volkov:1996br,Volkov:1996fk,Volkov:2006vq}).   
This condition can be written in the form 
\bea \label{I1f}
I_1^f = -i \frac{N_C}{(2\pi)^4} \int \frac{\dd^4k\; f({k^\bot}^2)}{m_u^2-k^2} = 0,
\eea
where $N_C=3$ is the number of colors. This gives gives $d\approx 1.78$~GeV$^{-2}$.
This means that the quark tadpole
connected with excited scalar mesons vanishes. Therefore the quark condensate acquires
only a contribution from the quark tadpole connected with the ground
scalar state~\cite{Volkov:1986zb}. 

After bosonizastion of the SU(2)$\times$SU(2) chiral symmetric 
four-quark Lagrangian and renormalization of the meson fields,
we get the following form of the quark--meson interaction: 
\bea
{\mathcal L}^{\mathrm{int}}&=& \bar{q}(k) \biggl\{ 
eQ\gamma_\mu A^\mu(p) +
\tau^3\gamma_5 \bigl[g_{\pi_1}\pi_1(p) + g_{\pi_2}\pi_2(p) f({k^\bot}^2)\bigr] 
\nonumber \\
&+& \frac{1}{2}\gamma_\mu\biggl[ g_{\rho_1}(\rho_1^\mu(p)\tau^3
+\omega_1^\mu(p)) 
+ g_{\rho_2}f({k^\bot}^2) (\rho_2^\mu(p)\tau^3 +\omega_2^\mu(p)) \biggr] 
\biggr\}q(k'). \qquad
\eea
Here $q$ are the $u$ and $d$ quark fields; $Q=diag(\frac{2}{3},-\frac{1}{3})$
is the quark charge matrix; $A^\mu$ is the electromagnetic field;
$\pi_i$, $\rho_i$, and $\omega_i$ are non-physical pseudoscalar and vector meson fields.
The coupling constants have the form~\cite{Volkov:1996br,Volkov:1996fk}:
\bea
&& g_{\pi_1} = \left[4I_2\biggl(1-\frac{6m_u^2}{M_{a_1}^2}\biggr)\right]^{-1/2},
\qquad
g_{\pi_2} = \left[4I_2^{f^2}\right]^{-1/2}, 
\nonumber \\ 
&& g_{\rho_1} = \left[\frac{2}{3}I_2\right]^{-1/2},
\qquad
g_{\rho_2} = \left[\frac{2}{3}I_2^{f^2}\right]^{-1/2}, 
\eea
where 
\bea
I_2 &=& - i N_c \int \frac{\dd^4k}{(2\pi)^4}
\frac{\Theta(\Lambda^2-\vec{k}^2)}{(m_u^2 - k^2)^2},   
\\ \nonumber
I_m^{f^n} &=& - i N_c \int \frac{\dd^4k}{(2\pi)^4}
\frac{\bigl(f({k^\bot}^2)\bigr)^n}{(m_u^2 - k^2)^m}, \quad n=1,2, \quad m=1,2.  
\eea
Note that in $g_{\pi_1}$ we take into account $\pi$---$a_1$ 
transitions~\cite{Volkov:1986zb},
$M_{a_1} = 1.23$~GeV is the $a_1$ meson mass. In the constant $g_{\pi_2}$ these 
transitions can be neglected, see Refs.~\cite{Volkov:1996br,Volkov:1996fk}. 

The free part of the Lagrangian for the pion fields contains
non-diagonal kinetic terms:
\bea
{\mathcal L}_{\pi}^{\mathrm{free}} = \frac{p^2}{2}
\biggl( \pi_1^2 + 2\Gamma_\pi\pi_1\pi_2 + \pi_2^2 \biggr)
- \frac{M_{\pi_1}^2}{2}\pi_1^2 - \frac{M_{\pi_2}^2}{2}\pi_2^2.
\eea
The mass terms have a diagonal form because of the condition~(\ref{I1f}), and
\bea
&& \Gamma_{\pi} = \frac{I_2^f}{\sqrt{I_2\, I_2^{f^2}}}\, ,
\nonumber \\
&& M_{\pi_1}^2 = g_{\pi_1}^2 \biggl[ \frac{1}{G_1} - 8I_1\biggr], 
\qquad
M_{\pi_2}^2 = g_{\pi_2}^2 \biggl[ \frac{1}{G_1 c_{\pi}} - 8I_1^{f^2}\biggr],
\eea
where $G_1=3.47$~GeV$^{-2}$ 
is the interaction constant of scalar and pseudoscalar quark currents in the initial
NJL model~\cite{Volkov:1996fk,Volkov:1997dd}.

The following transformation allows us to get
a diagonal form of the free meson Lagrangian~\cite{Volkov:1996fk,Volkov:1997dd}: 
\bea
&& \pi^0 = \pi_1\cos(\alpha-\alpha_0)-\pi_2\cos(\alpha+\alpha_0),
\nonumber \\
&& {\pi^0}' = \pi_1\sin(\alpha-\alpha_0)-\pi_2\sin(\alpha+\alpha_0),
\eea
where
\bea \label{alpha0}
\sin\alpha_0 = \sqrt{\frac{1+\Gamma_\pi}{2}}, 
\qquad
\tan(2\alpha - \pi) = \sqrt{\frac{1}{\Gamma_\pi^2}-1}
\left[\frac{M_{\pi_1}^2-M_{\pi_2}^2}{M_{\pi_1}^2+M_{\pi_2}^2}\right].
\eea
For the angles we obtained the following values $\alpha_0=59.06^\circ$ 
and $\alpha=59.38^\circ$.
The free pion Lagrangian takes the standard form
\bea
{\mathcal L}_{\pi}^{\mathrm{free}} = \frac{p^2}{2}\biggl( {\pi^0}^2 + {{\pi^0}'}^2 \biggr)
- \frac{M_\pi^2}{2}{\pi^0}^2 - \frac{M_{\pi'}^2}{2}{{\pi^0}'}^2.
\eea
For $c_\pi=1.36$ the values $M_\pi\approx 134.8$~MeV and $M_{\pi'}\approx 1308$~MeV were received,
in agreement with the experimental ones $134.9766\pm 0.0006$~MeV
and  $1300\pm 100$~MeV, respectively~\cite{Amsler:2008zzb}.

As the result the interaction Lagrangian for physical pion fields with quarks takes the form
\bea \label{L_int_pi}
{\mathcal L}^{\mathrm{int}}_{\pi}&=& \bar{q}(k) \tau^3\gamma_5 \biggl\{ 
 \bigg[ g_{\pi_1}\frac{\sin(\alpha+\alpha_0)}{\sin(2\alpha_0)}
       +g_{\pi_2}f({k^\bot}^2)\frac{\sin(\alpha-\alpha_0)}{\sin(2\alpha_0)}\biggr]\pi^0(p) 
\nonumber \\
&-&\bigg[ g_{\pi_1}\frac{\cos(\alpha+\alpha_0)}{\sin(2\alpha_0)}
        +g_{\pi_2}f({k^\bot}^2)\frac{\cos(\alpha-\alpha_0)}{\sin(2\alpha_0)}\biggr]{\pi^0}'(p) 
\biggr\}q(k').
\eea
An analogous procedure for the vector mesons leads to
\bea
{\mathcal L}^{\mathrm{int}}_{\rho,\omega}&=& \bar{q}(k) \frac{\gamma_\mu}{2} \biggl\{ 
 \bigg[ g_{\rho_1}\frac{\sin(\beta+\beta_0)}{\sin(2\beta_0)}
       +g_{\rho_2}f({k^\bot}^2)\frac{\sin(\beta-\beta_0)}{\sin(2\beta_0)}\biggr]
(\tau^3\rho^0_\mu(p)+\omega_\mu(p)) 
\nonumber \\
&-&\bigg[ g_{\rho_1}\frac{\cos(\beta+\beta_0)}{\sin(2\beta_0)}
        +g_{\rho_2}f({k^\bot}^2)\frac{\cos(\beta-\beta_0)}{\sin(2\beta_0)}\biggr]
(\tau^3{\rho^0}'_\mu(p)+\omega'_\mu(p)) 
\biggr\}q(k'), \qquad
\eea
where the angles  $\beta_0=61.53^\circ$ and $\beta=76.78^\circ$ are 
defined analogously to Eqs.~(\ref{alpha0}),
using 
\bea
M_{\rho_1}^2 = \frac{3}{8G_2I_2}\, ,\qquad
M_{\rho_2}^2 = \frac{3}{8c_\rho G_2I_2^{f^2}}\, .
\eea
For $c_\rho=1.15$ and $G_2 = 13.1$~GeV$^{-2}$
we get $M_\rho=M_\omega\approx 783$~MeV and
 $M_\rho'=M_\omega'=1450$~MeV. The corresponding 
experimental values are $M_{\rho}= 775.49 \pm 0.34$~MeV,
$M_{\omega} =782.65 \pm 0.12$~MeV,
$M_{\rho'} =1465 \pm 25$~MeV, and
$M_{\omega'}= 1425 \pm 25$~MeV.

Let us consider now the standard two-photon decay, described by the triangle quark diagram
of the anomalous type. The decay amplitude has the form
\bea
&& A^{\pi^0\to\gamma\gamma}_{\mu\nu} = 8m_u\varepsilon_{\mu\nu\gamma\sigma}q_1^{\gamma}q_2^{\sigma}
e^2\bigl( Q_u^2 - Q_d^2 \bigr) 
\nonumber \\ && \quad \times
\frac{(-i)N_C}{(2\pi)^4} 
\int\frac{\dd^4k}{(k^2-m_u^2+i0)((k-q_1)^2-m_u^2+i0)((k+q_2)^2-m_u^2+i0)}
\nonumber \\ && \quad \times 
\biggl\{
g_{\pi_1} \frac{\sin(\alpha+\alpha_0)}{\sin(2\alpha_0)}
+g_{\pi_2}f({k^\bot}^2)\frac{\sin(\alpha-\alpha_0)}{\sin(2\alpha_0)} 
\biggr\},
\eea
where $Q_{u,d}$ are the $u$ and $d$ quark charges; 
$q_{1,2}$ are the photon momenta.
The amplitude contains two types of one-loop integrals: 
with and without form factor in the quark-pion vertex.
The expression for the ${\pi^0}'\to \gamma\gamma$ amplitude
differs from the above one only by the coupling constant 
of pion and quarks, see Eq.~(\ref{L_int_pi}), and by the mass
of the decaying particle. In the calculation 
we take into account only the real part of the loop integrals.
This ansatz corresponds to the na\"ive 
confinement definition~\cite{Pervushin:1985yi}, which was used 
in some our recent works~\cite{Bystritskiy:2007wq,Bystritskiy:2009zz,Volkov:2009pc}.

Consider now two particles decay modes of pseudoscalar and vector mesons 
with a single photon. 
The amplitude of $\rho^0\to\pi^0\gamma$ takes the form
\bea
&& A^{\rho^0\to\pi^0\gamma}_{\mu\nu} = 4m_u\varepsilon_{\mu\nu\gamma\sigma}q_1^{\gamma}q_2^{\sigma}\;
2e\bigl( Q_u + Q_d \bigr) 
\nonumber \\ && \quad \times  
\frac{(-i)N_C}{(2\pi)^4} 
\int\frac{\dd^4k}{(k^2-m_u^2+i0)((k-q_1)^2-m_u^2+i0)((k+q_2)^2-m_u^2+i0)}
\nonumber \\ && \quad \times 
\biggl\{
\frac{g_{\rho_1}}{2} \frac{\sin(\beta+\beta_0)}{\sin(2\beta_0)}
+\frac{g_{\rho_2}}{2} f({k^\bot}^2)\frac{\sin(\beta-\beta_0)}{\sin(2\beta_0)} 
\biggr\}
\nonumber \\ && \quad \times 
\biggl(
g_{\pi_1} \frac{\sin(\alpha+\alpha_0)}{\sin(2\alpha_0)}
+g_{\pi_2}f({k^\bot}^2)\frac{\sin(\alpha-\alpha_0)}{\sin(2\alpha_0)} 
\biggr),
\eea
where $q_1$ and $q_2$ are the vector meson and photon momenta, respectively

First we recalculate the width of radiative decays of the ground meson
states, see Table~\ref{table:1}.
The results are in a satisfactory agreement with the
experimental data~\cite{Amsler:2008zzb}. 
Note that similar strong decays of considered here radial-excited mesons
$\rho'\to\omega\pi$ and $\omega'\to\rho\pi$ within the same non-local model 
were earlier found in Ref.~\cite{Volkov:1997dd} to be also 
in a satisfactory agreement with 
observations~\cite{Amsler:2008zzb,Clegg:1993mt}:
\bea
&& \Gamma_{\rho'\to\omega\pi}^{theor.} \approx 75\ \mathrm{MeV}, \qquad
\Gamma_{\rho'\to\omega\pi}^{exper.} = 65.1\pm 12.6\ \mathrm{MeV}, \\
&& \Gamma_{\omega'\to\rho\pi}^{theor.} \approx 225\ \mathrm{MeV}, \qquad
\Gamma_{\omega'\to\rho\pi}^{exper.} = 174\pm 60\ \mathrm{MeV}. 
\eea

We remind that the decay width of the ground states have been obtained
within the local NJL model in 1986~\cite{Volkov:1986zb} 
in a good agreement with the experimental data. 
In the considered here non-local version of the NJL model radial--excited meson states 
are mixed with the ground ones. However this mixing
does not lead to distortion of the description of the ground meson states interaction 
among each other, received in the local model.

\begin{table}[ht]
\caption{Ground state meson radiative decays widths.
\label{table:1}}
\bigskip
\begin{tabular}{|c|c|c|c|}
\hline
Decay          
& ${\pi^0}\to\gamma\gamma$ 
& ${\rho^0}\to\pi^0\gamma$  
& ${\omega^0}\to\pi^0\gamma$
\\
\hline
Theory 
& 7.7 eV 
& 77 keV 
& 710 keV 
\\
\hline
Exper. 
& 7.5 $\pm$ 1.1 eV
& 88  $\pm$ 12 keV
& 700 $\pm$ 30 keV 
\\
\hline
\end{tabular}
\end{table}

\begin{table}[ht]
\caption{Radiative decay widths of ${\pi^0}'$ and ${\rho^0}'$
\label{table:2}}
\bigskip
\begin{tabular}{|c|c|c|c|c|}
\hline
Decay          
& ${\pi^0}'\to\gamma\gamma$ 
& ${\pi^0}'\to\rho^0\gamma$ 
& ${\rho^0}'\to\pi^0\gamma$ 
&  ${\rho^0}'\to{\pi^0}'\gamma$ 
\\
\hline
Theory 
& 3.2~keV 
& 1.8 keV 
& 450 keV 
& 24 keV
\\
\hline
\end{tabular}
\end{table}

\begin{table}[ht]
\caption{Decay widths of radiative processes with $\omega$ and $\omega'$ mesons.
\label{table:3}}
\bigskip
\begin{tabular}{|c|c|c|c|}
\hline
Decay           
& ${\pi^0}'\to\omega\gamma$ 
& ${\omega}'\to\pi^0\gamma$ 
& ${\omega}'\to{\pi^0}'\gamma$ 
\\
\hline
Theory 
& 17 keV 
& 3.7 MeV 
& 93 keV 
\\
\hline
\end{tabular}
\end{table}

Tables~\ref{table:2} and \ref{table:3} contain our results of theoretical calculations
of radiative decays widths of the  radial--excited states of
$\pi^0$, $\rho^0$ and $\omega$ mesons calculated in frames of the NJL model.
For the calculations of the phase space we used the present 
experimental values for the excited meson masses.
The widths of these decay 
channels are not yet measured experimentally. So we made
predictions which can be relevant for future experiments.

We plan to perform similar calculations to describe radiative decay 
channels of radial--excited of $\eta$, $\eta'$, and $\phi$ mesons 
in non-local U(3)$\times$U(3) NJL model and, besides, to investigate production
of the radial--excited mesons at modern $e^+e^-$ colliders.

\subsection*{Acknowledgments}
The authors are grateful to L.I.~Volkova for the help in manuscript preparation. 
We also are grateful to A.~Akhmedov for discussions. 
The work was supported by the RFBR grant 10-02-01295-a.

\end{document}